\documentclass[twocolumn,aps,prl,groupedaddress,showpacs]{revtex4}
\usepackage{epsfig,amssymb,amsmath}
\def\sH{\mathcal{H}}\def\sT{\mathcal{T}}
\def\Tr{\mbox{Tr}}\def\map#1{\mathcal{#1}}

\begin{document}
\title{Quantum calibration of measuring apparatuses} \author{Giacomo Mauro
  D'Ariano}
\email{dariano@unipv.it} 
\altaffiliation[Also at ]{Center for Photonic Communication and Computing, Department of
  Electrical and Computer Engineering, Northwestern University,
  Evanston, IL 60208}
\author{Lorenzo Maccone}\email{maccone@unipv.it} 
\author{Paoloplacido Lo Presti}\email{lopresti@unipv.it} 
\affiliation{QUIT - Quantum Information
  Theory Group\homepage{http://www.qubit.it}, Dipartimento di Fisica ``A.  Volta'' Universit\`a di Pavia, via A.
  Bassi 6, I-27100 Pavia, Italy.}
\date{\today}
\begin{abstract}
By quantum calibration we name an experimental procedure apt to completely characterize an unknown 
measurement apparatus by comparing it with other calibrated apparatuses. Here we show how to achieve
the calibration of an arbitrary measuring apparatus, by using it in conjunction with a ``tomographer''
in a correlation setup with an input bipartite system.  The method is robust to imperfections of the 
tomographer, and works for practically any input state of the bipartite system.  
\end{abstract}
 \maketitle 

The calibration of measuring apparatuses is at the basis of any experiment. Theory and experiment are
unavoidably interwoven, and the calibration procedure often needs a detailed knowledge of the inner
working of the apparatus, especially at extreme precisions and sensitivities, where a quantum mechanical
description is needed. Here,  the actual ``observable'' that is measured depends crucially
on the microscopic details of the apparatus, and without knowing them the measurement lacks 
physical interpretation. 
\par In a quantum mechanical description, the calibration of a measuring apparatus corresponds to
the knowledge of its POVM (positive operator-valued measure~\cite{Helstrom}), which gives the
probability $p(n)$ of any measurement outcome $n$ for arbitrary input state, via the Born rule 
\begin{equation}
p(n)=\Tr[\rho P_n].\label{povm}
\end{equation}
In Eq. (\ref{povm}) $\rho$ is the density operator of the state on the Hilbert space $\sH$ of the 
system, and the POVM is given by the set of operators $\{P_n\}$ on $\sH$. To ensure
that $p(n)$ is a probability, the POVM must satisfy the positivity and normalization constraints $P_n\geqslant 0$, $\sum_nP_n=I$.
\par The concept of POVM generalizes the familiar von Neumann observable describing perfect
measurements. Here the probability of obtaining outcome $n$ is given by
$p(n)=\left|\langle\psi|o_n\rangle\right|^2$, $\{|o_n\rangle\}$ denoting a complete orthonormal  
basis for $\sH$, i. e. with POVM given by the one-dimensional projectors
$P_n=|o_n\rangle\langle o_n|$. The physical interpretation of the measurement is 
given via a quantization rule that associates a self-adjoint operator $O$ to a classical 
observable, $|o_n\rangle$ being the eigenvector of $O$ corresponding to its $n$th eigenvalue
$o_n$. However, this concept of observable does not cover many practical situations---e.~g.
phase-estimation\cite{physcri,pomph},  joint measurements of incompatible observables~\cite{Art,GOLU},
discrimination among non-orthogonal states~\cite{yuenlax,chefles}, informationally complete
measurements~\cite{univest}, transmission of reference frames~\cite{refframe}---and here the POVM
description is needed. But then, in absence of a direct physical interpretation of the measurement, 
we are faced with the problem of assessing the correct functioning of the measuring apparatus. 
\par Inferring the POVM of an apparatus through the theoretical description of its functioning
leads to quite involved derivations, based on different kinds of approximations. 
A paradigmatic case is that of  the photo-counter~\cite{Mandelbook}, where the number of photons
claimed to be detected---usually very uncertain---is typically inferred from the cascading mechanism
of the amplification process. The calibration is given essentially in terms of quantum efficiency
and  dark-current, and mostly saturation effects categorize detectors into the major classes of
``linear'' and ``single-photon''. Even in a very simplified model, a theoretical description
accounting  for the above features is very involved~\cite{KK,Mandel}, and the resulting theoretical
calibration is exceedingly indirect.  
\par
\begin{figure}[hbt]
\begin{center}
\epsfxsize=.6
\hsize\leavevmode\epsffile{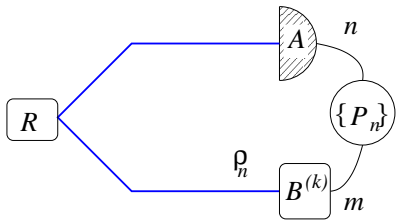}
\end{center}
\caption{Experimental setup to determine the POVM of the unknown measurement apparatus~A. The
apparatus A is used jointly with a ``tomographer'' on a bipartite 
system prepared in a pre-determined state~$R$. The tomographer measures an observable $B^{(k)}$ 
from the quorum $\{B^{(k)}\}$, yielding result $m$, whereas the unknown apparatus gives outcome
$n$. The joint outcomes $(n,m)$ are then processed using a tomographic algorithm, to finally obtain
the POVM $\{P_n\}$ of~A. } 
\label{f:experiment}\end{figure}

 The above scenario raises the following problem: is it possible to calibrate a measuring 
apparatus---i. e. to determine its POVM---with a purely experimental procedure, e.~g. by comparing the
apparatus with other (previously calibrated) apparatuses? In this paper we propose a method to
determine a POVM experimentally. The method uses the unknown apparatus jointly with a calibrated
``tomographer'' on a suitably prepared bipartite system, as in Fig.~\ref{f:experiment}, and the
calibration results from the analysis of the correlations of outcomes.
 [A tomographer is an apparatus that measures an observable tunable in a complete set called {\em
 quorum}: more details on quantum tomography will be given in the following]. The basic scheme of the method stems on a
 previous method for the tomographic  
reconstruction of quantum operations~\cite{qop1}, and generalizes a popular calibration
scheme~\cite{klishko,sergienko} designed to determine the quantum efficiency of a
photo-detector. As it will be shown in the following, there is ample freedom in the choice of both
the input bipartite state and the tomographer. The joint measurement must be repeated many times, 
analyzing the measurement outcomes with a proper tomographic algorithm~\cite{tomog,maxlik}, 
the POVM calibration being approached in the limit of infinitely many outcomes. For finite set of
data, the reconstructed POVM will be affected by statistical errors, which can be precisely
estimated via the tomographic algorithm. The method works for generally infinite-dimensional Hilbert 
space (yielding a finite number of POVM elements, corresponding to the actually occurred outcomes).

The following simple example illustrates how the procedure works.
Suppose we know that the apparatus measures an observable, but we don't know which one, and denote
it by the orthonormal basis $\{|o_n\rangle\}$. We can use the maximally entangled input state
$|\Psi\rangle=\sum_{i=1}^d|i\rangle|i\rangle/\sqrt{d}$ in the space $\sH\otimes\sT$, $\sT$
denoting the space of the quantum system impinging into the tomographer. 
The state can be equivalently written as $|\Psi\rangle=
\frac 1{\sqrt{d}}\sum_{j=1}^d|o_j\rangle|o_j^*\rangle$ ($|o_j^*\rangle$ denotes the vector with 
the complex conjugated coefficients of $|o_j\rangle$ with respect to the basis
$\{|i\rangle\}$). Then, the outcome $n$ of the unknown measuring apparatus conditions the state
$\rho_n=|o_n^*\rangle\langle o_n^*|$ at the tomographer, and the POVM can be recovered  
using state reconstruction.  
\medskip
\par We now present the general quantum calibration procedure. Let's fix one observable at the
tomographer, and denote it by $\{|b_n\rangle\}$. Upon denoting by $\{P_n\}$
the POVM of our measuring apparatus that we want to calibrate, the Born rule (\ref{povm}) predicts
that the outcome $(n,m)$ of the joint measurement will occur with probability
\begin{equation}
p(n,m)=\Tr\left[(P_n\otimes|b_m\rangle\langle b_m|)R\right],
\label{born}
\end{equation}
where $R$ is the joint state of the two quantum systems, and we remind that the POVM of the joint
measurement is given by the tensor product of the individual POVM's. Upon rewriting the joint 
probability in terms of the conditional probability $p(m|n)$ via Bayes' rule, we conveniently
introduce the state $\rho_n$ at the tomographer conditioned by outcome $n$ at the unknown 
measuring apparatus, namely
\begin{equation}
p(n,m)\doteq p(n)\;p(m|n)\doteq
p(n)\Tr\left[\rho_n|b_m\rangle\langle b_m|\right]. 
\label{born1}
\end{equation}
Upon evaluating the trace in Eq.~(\ref{born}) in two steps, i.~e.
\begin{equation}
p(n,m)=\Tr\Big[|b_m\rangle\langle
b_m|\Tr_1\left[\left(P_n\otimes\openone\right)R\right]\Big],
\label{pp}
\end{equation}
and by equating Eqs. (\ref{born1}) and (\ref{pp}) for any possible vector $|b_m\rangle$
(i. e. any possible observable), we have $\rho_n p(n)=\Tr_1[(P_n\otimes\openone)R]$, namely
\begin{equation}
\rho_n=\frac{\Tr_1[(P_n\otimes\openone)R]}{\Tr[(P_n\otimes\openone)R]},\quad
p(n)=\Tr[(P_n\otimes\openone)R].
\label{pp1bis}
\end{equation}
The POVM element $P_n$ can be recovered from the conditioned state $\rho_n$ as follows
\begin{equation}
P_n=p(n)\map{R}^{-1}(\rho_n),
\label{pin}
\end{equation}
by inverting the map 
\begin{equation}
\map{R}(X)\doteq\Tr_1[(X\otimes\openone)R],
\end{equation}
$X$ denoting an operator on $\sH$. The map $\map{R}$ depends only on the input state 
$R$, which then must be known. Hence we need a {\em pre-calibration} stage in which we
previously determine the joint state $R$ (this can be done via a joint quantum tomography with two
equal tomographers on the input state $R$). Invertibility of the map $\map{R}$ corresponds to a
so-called {\em faithful state}~\cite{qop1}. Since invertible maps are a dense set, then almost 
any quantum state $R$ is faithful. Of course, when approaching a state corresponding to a non
invertible map, some information on the POVM $\{P_n\}$ will be lost, corresponding to
increasingly large statistical errors for some matrix elements of the operators $P_n$ (inverting a
linear map is clearly equivalent to inverting an operator: for the reader who prefers operators to map,
an explicit connection between operators and maps is given in Ref.~\cite{qop1}). 
\par Once the inverse map $\map{R}^{-1}$ has been calculated, we use quantum tomography in order to
recover $\rho_n$. Shortly quantum  tomography is a method that allows us to estimate the ensemble
average of an arbitrary (complex) operator $X$ by measuring a set of observables $\{B^{(k)}\}$, called
{\it quorum}, which span the space of operators of the system (for recent reviews on quantum
tomography, see Refs.~\cite{tomog,tomobook}). Typical examples of quorums are the three Pauli matrices
$\sigma_x$, $\sigma_y$, $\sigma_z$ for a qubit, or the set of 
quadratures $X_\phi=\frac{1}{2}(a^\dag e^{i\phi}+a e^{-i\phi})$ for a single mode of the
radiation field with annihilation and creation operators $a$ and $a^\dag$, respectively,
$\phi\in[0,\pi)$ playing the role of the observable label within the quorum $\{X_\phi\}$
($X_\phi$ is measured by a homodyne detector at phase $\phi$ relative to the local
oscillator~\cite{tomog}). In short, the generic operator $X$ is expanded as
$X=\sum_k\Tr[XC^{(k)}{}^\dag]B^{(k)}$, $\{C^{(k)}\}$ denoting a dual set of the quorum $\{B^{(k)}\}$  
(the sum being replaced by an integral for continuous $k$), the quantity $\Tr[X C^{(k)}{}^\dag]$ being
evaluated analytically. Notice that one can remove the effects of noise at the tomographer
if the noise map $\map{N}$ is invertible, by writing $X=\sum_k\Tr[{\cal
  N}^{-1}(X)C^{(k)}{}^\dag]{\map{N}}(B^{(k)})$. 
\par For the tomographic reconstruction we can either: {\em a)} average over the
quorum, e.~g.  estimate $\langle X\rangle$ via the ensemble averages of the quorum observable
as $\langle X\rangle=\sum_k\Tr[XC^{(k)}{}^\dag]\langle B^{(k)}\rangle$ (the estimation of the density 
matrix element $\rho_{ij}$ corresponding to $X=|j\rangle\langle i|$); {\em b)} we can use the
maximum likelihood approach~\cite{maxlik}. In this case, the estimated POVM elements $P_n$ will
maximize the probability $\Tr[(P_n\otimes|b^{(k)}_m\rangle\langle b^{(k)}_m|)R]$ in the joint
measurement on the pre-determined input state $R$ of getting outcome $n$ on the unknown measuring
apparatus and $m$ for the $k$th observable $B^{(k)}$ of the quorum. Equivalently, one can maximize the
logarithm of this quantity and consider simultaneously all the $N$ joint measurement outcomes,
corresponding to maximizing the {\em likelihood functional} 
\begin{equation}
{\cal L}(\{P_n\})\equiv\sum_{i=1}^N\log\Tr\Big[(P_{n_i}\otimes
|b^{(k_i)}_{m_i}\rangle\langle b^{(k_i)}_{m_i}|)R\Big], 
\label{mlik}
\end{equation}
under the constraints $P_n\geqslant 0$ and $\sum_nP_n=I$. Other prior knowledge about $P_n$
can be easily incorporated by adding further constraints. Moreover, we can account for a known
source of noise $\map{N}$ at the tomographer, by replacing the projector $|b^{(k_i)}_{m_i}\rangle\langle
b^{(k_i)}_{m_i}|$ in Eq.~(\ref{mlik}) with $\map{N}(|b^{(k_i)}_{m_i}\rangle
\langle b^{(k_i)}_{m_i}|)$. 
\medskip
\par Therefore, the procedure to calibrate an unknown measurement apparatus can be summarized in
the following steps: {\em i)}~[{\em pre-calibration}] using two tomographers, reconstruct the input
joint state $R$. Check whether $R$ is faithful; {\em ii)}~[{\em joint measurements with the unknown
apparatus}] Replace one tomographer with the unknown detector, and collect $N$ pairs of outcomes
$\{n_i,m_i\}$, $i=1,...,N$ in a set of joint measurements with randomly selected observable
$B^{(k_i)}$ in the quorum; {\em iii)}~[{\em Data analysis}] From the
experimental data collect the probability $p(n)$ of the outcome $n$ at the unknown measurement
apparatus, and then estimate the POVM $\{P_n\}$ using a tomographic strategy---either the 
averaging or the maximum likelihood. In the first case evaluate the density matrix $\rho_n$ of the
state impinging in the unknown measuring apparatus, and then use Eq. (\ref{pin}) to recover the
POVM. In the second case, evaluate the POVM directly by maximizing the likelihood functional $\cal L$ in
Eq.~(\ref{mlik}) on the given set of experimental data, with the state $R$ obtained at 
step~{\em i)}. 
\begin{figure}[h!]
\begin{center}
\epsfxsize=1.
\hsize\leavevmode\epsffile{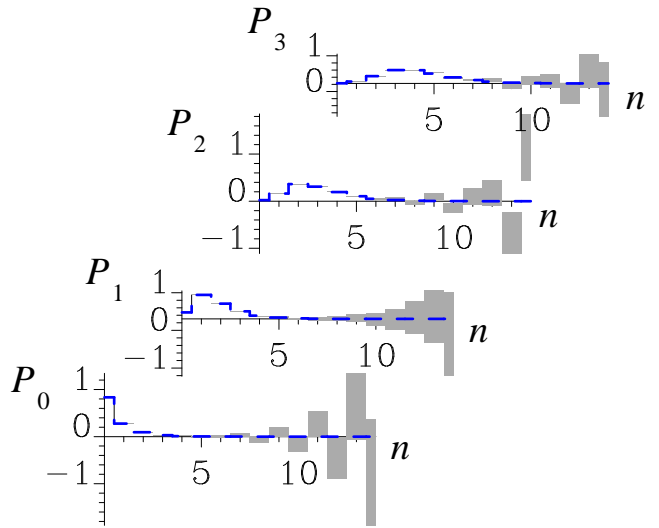}
\end{center}
\caption{Calibration of the photodetector in Fig. ~\ref{f:detector} with $\eta_p=80\%$ and $\nu=1$,
using a twin-beam with $\xi=0.88$ (see text), and homodyne tomography with quantum efficiency
$\eta_h=90\%$. The plots are the reconstruction of the diagonal matrix elements $\langle
n|P_k|n\rangle$ of the detector POVM, from a set of $5\times 10^6$ compute-simulated data, using 
the averaging strategy. The reconstructed POVM is at the middle of the error-bars, whereas the
theoretical values, for comparison, are given by the thick line.} 
\label{f:tomog}\end{figure} 
\begin{figure}[htb]
\begin{center}
\epsfxsize=.6
\hsize\leavevmode\epsffile{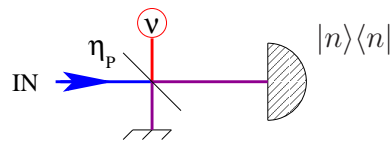}
\end{center}
\caption{Model of the photodetector calibrated in the simulations of Figs.~\ref{f:tomog}
and~\ref{f:maxlik}. Non-unit quantum efficiency $\eta_p$ and dark-current with averaged
photon number $\nu$ are equivalent to preceding an ideal detector by a beam-splitter of transmissivity
$\eta_p$ mixing the input signal with a thermal mode with $\nu$ average photons.}
\label{f:detector}\end{figure}
\par In Fig. \ref{f:tomog} we present a simulated experiment of the quantum calibration of a
photo-counter using homodyne tomography with the averaging strategy. 
The model of the calibrated detector is given in Fig. \ref{f:detector}.
Since the resulting POVM is diagonal in the photon-number
basis, we limit the reconstruction to the diagonal elements only. As input state $R$ we use a twin
beam state from parametric down-conversion of vacuum, of the form
$\propto\sum_m\xi^m|m\rangle\otimes|m\rangle$, 
where $\xi$ is related to the amplification gain and $|m\rangle$ denotes the eigenstate of the
photon number. One can easily check that the twin beam is faithful for all $\xi\neq 0$. As typical
imperfection of the tomographer, we consider non-unit quantum efficiency $\eta_h$ for the homodyne
detector (the noise map can be inverted as long as $\eta_h>\frac{1}{2}$~\cite{tomog}).  Since we
reconstruct only the diagonal part of the POVM, one can easily show that there is non need of
knowing the homodyne phase $\phi$, which, however, must be randomly distributed (the knowledge of
$\phi$ would  allow to recover also the off-diagonal elements of the POVM).
\begin{figure}[htb]
\begin{center}
\epsfxsize=1.
\hsize\leavevmode\epsffile{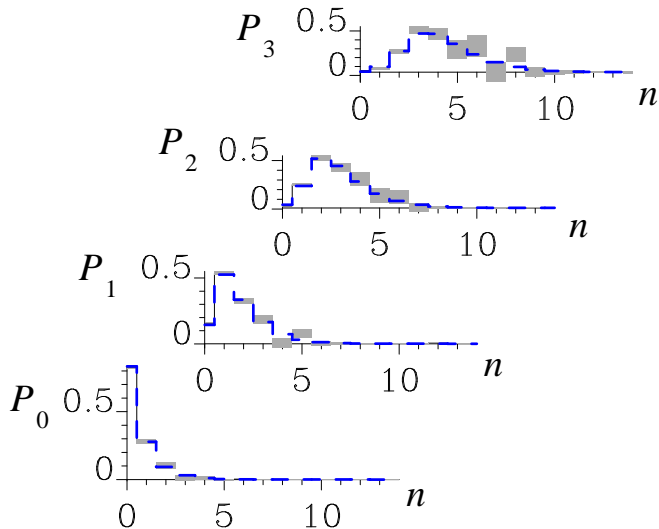}
\end{center}
\caption{The same as in Fig.~\ref{f:tomog}, but using the maximum likelihood method.
Here only $5\times 10^4$ simulated data are used. The error bars are obtained by 
standard bootstrapping techniques over a virtual repetition of $50$ experiments. Notice how the
result is statistically less noisy than that in Fig.~\ref{f:tomog}, even for a $10^{-2}$ smaller
set of data.}  
\label{f:maxlik}\end{figure} 
\par In Fig. \ref{f:maxlik} we present the same calibration, but using the maximum likelihood
strategy.  The convergence of the maximum-search algorithm is assured by the strict convexity of the
likelihood functional $\cal L$ over the space of diagonal POVM's (the convergence speed, however, can
be practically very slow). In the simulation we used a blend of sequential quadratic programming
routines (to perform the constrained maximization) along with expectation-maximization
techniques~\cite{maxlik}. 
By comparing Figs. \ref{f:tomog} and \ref{f:maxlik} we can see how the maximum likelihood 
estimation is more statistically efficient (i.~e. fewer data are needed to  achieve 
the same statistical error) than the averaging strategy~\cite{notelik}, and, in addition,
the maximization of the likelihood recovers all the POVM elements simultaneously.
On the other hand, compared to the averaging strategy, the maximum likelihood approach has the
drawback of being biased, since one needs to put a cut-off to the Hilbert space dimension of the
tomographic reconstruction and/or to the cardinality of the POVM. Both simulated experiments use
realistic parameters and are feasible in the lab with current technology (see, for example,
Refs.~\cite{dakum,lvovsky,grangier,zavatta}), the major challenge of a real experiment being the
matching of modes between photo-counter and homodyne detector, also ensuring that the detected modes
are the same of the pre-calibration stage.  

\par We acknowledge financial support by INFM PRA-2002-CLON and MIUR for Cofinanziamento 2003.  
PL acknowledges partial support from ATESIT project IST-2000-29681.


\begin{references}
\bibitem{notelik} This is true in general on theoretical-statistics grounds, since when the optimal
estimator (i.e. the one achieving the Cramer-Rao bound) exists, then it coincides with the maximum
likelihood estimator.
\end{references}
\end{document}